%% file: Manuscript.tex
\documentclass[
footinbib,fleqn,aps,prb,
twocolumn,superscriptaddress,reprint,
]{revtex4-2}
\usepackage[]{amsmath}
\usepackage{amssymb}
\usepackage{bm}
\usepackage{mathtools}
\usepackage[dvips]{graphicx}
\usepackage{color}
\usepackage{tabularx}
\usepackage{threeparttable}
\usepackage{multirow}
\usepackage{adjustbox}
\usepackage{booktabs}           % table ruler
\usepackage{makecell}           % table vertical ruler
\usepackage{hyperref}           % hyper reference
\usepackage[version=4]{mhchem}  % chemical
\usepackage{siunitx}
\usepackage{soul}
\usepackage{subfiles}



\begin{document}

\title{Deep learning generative model for crystal structure prediction}

\author{Xiaoshan Luo}
\affiliation{Key Laboratory of Material Simulation Methods and Software of Ministry of Education, College of Physics, Jilin University, Changchun 130012, P. R. China}
\affiliation{State Key Laboratory of Superhard Materials, College of Physics, Jilin University, Changchun 130012, P. R. China}

\author{Zhenyu Wang}
\affiliation{Key Laboratory of Material Simulation Methods and Software of Ministry of Education, College of Physics, Jilin University, Changchun 130012, P. R. China}
\affiliation{International Center of Future Science, Jilin University, Changchun, 130012, P. R. China}

\author{Pengyue Gao}
\affiliation{Key Laboratory of Material Simulation Methods and Software of Ministry of Education, College of Physics, Jilin University, Changchun 130012, P. R. China}

\author{Jian Lv}
\email{lvjian@jlu.edu.cn}
\affiliation{Key Laboratory of Material Simulation Methods and Software of Ministry of Education, College of Physics, Jilin University, Changchun 130012, P. R. China}

\author{Yanchao Wang}
\email{wyc@calypso.cn}
\affiliation{Key Laboratory of Material Simulation Methods and Software of Ministry of Education, College of Physics, Jilin University, Changchun 130012, P. R. China}

\author{Changfeng Chen}
\email{chen@physics.unlv.edu}
\affiliation{Department of Physics and Astronomy, University of Nevada, Las Vegas, NV 89154, USA}

\author{Yanming Ma}
\email{mym@jlu.edu.cn}
\affiliation{Key Laboratory of Material Simulation Methods and Software of Ministry of Education, College of Physics, Jilin University, Changchun 130012, P. R. China}
\affiliation{International Center of Future Science, Jilin University, Changchun, 130012, P. R. China}

\date{\today}

\begin{abstract}
Recent advances in deep learning generative models (GMs) have created  high capabilities in accessing and assessing complex high-dimensional data, allowing superior efficiency in navigating vast material configuration space in search of viable structures. Coupling such capabilities with physically significant data to construct trained models for materials discovery is crucial to moving this emerging field forward. Here, we present a universal GM for crystal structure prediction (CSP) via a conditional crystal diffusion variational autoencoder (Cond-CDVAE) approach, which is tailored to allow user-defined material and physical parameters such as composition and pressure. This model is trained on an expansive dataset containing over 670,000 local minimum structures, including a rich spectrum of high-pressure structures, along with ambient-pressure structures in Materials Project database. We demonstrate that the Cond-CDVAE model can generate physically plausible structures with high fidelity under diverse pressure conditions without necessitating local optimization, accurately predicting 59.3\% of the 3,547 unseen ambient-pressure experimental structures within 800 structure samplings, with the accuracy rate climbing to 83.2\% for structures comprising fewer than 20 atoms per unit cell. These results meet or exceed those achieved via conventional CSP methods based on global optimization. The present findings showcase substantial potential of GMs in the realm of CSP.
\end{abstract}

\maketitle

\section*{Introduction}
Crystal structure fundamentally shapes all physical and chemical properties of materials, making the determination of atomic arrangements that define the crystal structure an essential and crucial task across a wide range of scientific disciplines \cite{Olson.review.2000}. Crystal structure prediction (CSP) via sound theoretical strategy and efficient computational algorithm plays a prominent role in state-of-the-art materials research by accelerating the discovery of desired functional materials compared to the traditional high-cost and low-efficiency trial-and-error attempts \cite{Wang.PerspectiveCSP.2014, Oganov.review.2019, Wang.review.2022}. CSP is rooted in the principles of thermodynamics, primarily tasked to identify the crystal structure with the lowest thermodynamic potential. Setting aside temperature and pressure effects that may be addressed via additional technical procedures, this task typically amounts to global optimization on the high-dimensional potential energy surface (PES). Over the past two decades, a plethora of methods/software have been developed to tackle the highly challenging CSP problem. Exemplary among them are AIRSS~\cite{Pickard.AIRSS.2011}, Basin Hopping~\cite{Oakley.BasinHopping.2013}, CALYPSO~\cite{Wang.CALYPSO.2010, Wang.CALYPSO.2012, Shao.Symmetry-CALYPSO.JCP.2022}, Minima Hopping~\cite{Goedecker.MinimaHopping.2004}, and USPEX~\cite{Glass.USPEX.2006, Laykhov.USPEX.2013}. The majority of these methods combine first-principle-based structural local optimization with sophisticated PES sampling techniques such as random sampling, molecular dynamics, evolutionary algorithms, and swarm intelligence. These methods have now become a staple in modern condensed matter physics and materials science, leading to exciting theory-driven breakthroughs, such as the discovery of high-pressure superhydride superconductors with record-setting critical temperatures~\cite{Wang.CaH6.2012, Peng.RE-H.PRL.2017, Liu.PNAS.2017, Drozdov.LaH10.Nature.2019, Somayazulu.PRL.2019}. Despite considerable and widespread successes, the NP-hard nature of the CSP problem leads to poor scaling performance with respect to the number of atoms or types of species in the material, which significantly impedes the application of CSP to complex systems. Consequently, developing more efficient and reliable methods for CSP continues to be a relentless pursuit in this field. Recently, a formal solution to the CSP problem was proposed that integrates combinatorial and continuous optimization \cite{Gusev.CombOpt.2023}, but its practical application hinges on quantum computing, which remains on the horizon and is not yet within immediate reach.

Meanwhile, deep learning, an emerging and rapidly evolving data-driven approach initially devised for handling images and natural languages, offers exceptional capabilities in representation and inference tasks for diverse data types, which is starting to make a major impact on materials research. The latest progress in supervised learning leads to the emergence of a series of methods for predicting various material properties, which significantly expands the scale of atomic simulations and accelerates computational materials discovery~\cite{Butler.review.2018, Schmidt.review.2019, Xie.CGCNN.2018, Chen.M3GNet.2022, Batzner.NequIP.2022, Deng.CHGNet.2023, Zhang.DPA-2.2023, Batatia.MACE-MP-0.2023, Merchant.GNoME.2023}. Moreover, the development of unsupervised learning for generative models (GMs) offers a promising alternative to sample the configuration space of materials~\cite{Sanchez-Lengeling.review.2018, Yan.review.2023}. Observations reveal that locally stable structures are mostly distributed across low-dimensional manifolds within the expanse of high-dimensional configuration space~\cite{Shires.SHEAP.2021}. In this scenario, generative models such as Variational Autoencoder (VAE)~\cite{Kingma.VAE.2013}, Generative Adversarial Network (GAN)~\cite{Goodfellow.GAN.2014}, diffusion model~\cite{Sohl-Dickstein.DiffusionModel.2015}, and flow-based model~\cite{Dinh.NICE.2014} serve as optimal tools for learning about the intrinsic distribution from physically meaningful structures (be they locally or globally stable) in well-curated structure databases. For example, VAE consists of an encoder that compresses data into a latent space with a simple probability distribution and a decoder that reconstructs data from the latent space, while the GAN learns about data distribution implicitly through a competitive interaction between a generator and a discriminator. After this learning process, the resulting models can generate plausible structures according to the learned data distribution, thereby opening a  promising avenue to address the complexities of the CSP problem.

The application of GMs to inorganic crystals lags behind that of protein~\cite{Huang.review.2016, Kuhlman.review.2019, Ingraham.Chroma.2023} and molecules~\cite{Sanchez-Lengeling.review.2018, Gebauer.cG-SchNet.2022}, but has started to catch up recently. iMatGen~\cite{Noh.iMatGen.2019} utilized a VAE to generate crystal structures of the V-O system, where a 3D voxel-based representation for structures was proposed as model input, while CrystalGAN~\cite{Nouira.CrystalGAN.2018} utilized a GAN and 2D coordinate-based representation to generate ternary hydrides from training sets of binary hydrides. Both models demonstrate the potential of GMs to produce plausible  crystal structures. Subsequent efforts have focused on enhancing the representation of structures~\cite{Hoffmann.2019, Kim.ZeoGAN.2020, Zhao.CubicGAN.2021, Lee.RDF-VAE.2021, Ren.FTCP.2022, Xie.CDVAE.2022, Xiao.SLI2Cry.NC.2023, Yang.UniMat.2023, Alverson.CrysTens.2024}, refining the model architecture~\cite{Court.Cond-DFC-VAE.2020, Kim.wGAN.2020, Xie.CDVAE.2022, Zeni.MatterGen.2023}, incorporating physical constraints~\cite{Zhao.PGCGM.2023, AI4Science.Crystal-GFN.2023, Jiao.DiffCSP.2023, Liu.PCVAE.2023, Jiao.DiffCSP++.2024}, and enabling conditional generation for desired compositions or properties~\cite{Long.DCGAN.2021, Han.LatentSpaceCDVAE.2023, Yang.UniMat.2023}. As a result, the chemical space accessible to a GM is gradually expanding with improving efficiency and quality of configuration sampling. Recently, a Crystal Diffusion VAE (CDVAE)~\cite{Xie.CDVAE.2022} was developed, in which the decoder adopts a noise conditional score network that relaxes atoms in crystals through Langevin Dynamics. Powered by SE(3) equivariant message-passing neural networks~\cite{Gasteiger.DimeNet.2020, Gasteiger.DimeNet++.2020, Gasteiger.GemNet.2021}, the CDVAE enables the consideration of key crystal attributes such as the invariance under permutation, translation, rotation, and periodicity. It demonstrates superior performance in generating valid, diverse, and realistic materials. In a similar vein, recent advances based on equivariant message-passing neural networks and diffusion models have led to GMs that process lattice and atom positions at the same time, further improving the quality of the generated structures~\cite{Zeni.MatterGen.2023, Jiao.DiffCSP.2023, Jiao.DiffCSP++.2024}. In parallel, the utilization of autoregressive large language models for crystal generation has also proven to be a promising approach in this field. Notably, the integration of physical biases, such as space group symmetry, into autoregressive models has been achieved, which greatly simplifies the task\cite{Antunes_CrystalLLM_2023, Gruver_CrystalLLM_2024, Cao_CrystalFormer_2024}.

Although GMs have shown great potential for efficient sampling of configuration space, thereby facilitating material discovery, their application to CSP is still in its early stages. Some GMs are specialized for particular chemical systems or lattice types, while others encompass a wider chemical space but are limited to generating structures under physically limited conditions, e.g., at zero pressure. In materials research, there is often a need to determine the ground-state structure of a specific chemical system under a given external pressure, which is crucial for mapping out the pressure-temperature phase diagrams, and for identifying promising high-pressure structures with desired properties, such as structures with exotic chemistry or high-temperature superconductivity~\cite{Wang.PerspectiveCSP.2014, Zhang.review.2017}. Therefore, there is an urgent need to develop a universal GM that is capable of generating structures with user-specified key physical conditions such as composition and pressure for CSP tasks. This necessitates the development of a conditional GM and accompanying databases that can capture sufficient information about structures at high pressure.

In this work, we have curated a comprehensive dataset containing 670,979 locally stable structures encompassing 86 elements and spanning a broad pressure range, which came from previously untapped high-pressure structure data collected from CSP simulations done by the wide CALYPSO community, and along with data from Materials Project (MP)~\cite{Jain.MaterialsProject.2013}, the combined results are henceforth termed the MP60-CALYPSO dataset. We devised a conditional GM for CSP based on CDVAE, termed Cond-CDVAE. By training the Cond-CDVAE model on the MP60-CALYPSO dataset, we constructed a universal GM capable of generating valid, diverse crystal structures conditioned on user-defined chemical composition and pressure, making it particularly suitable for pertinent CSP tasks. Our extensive benchmarking of the Cond-CDVAE model showcases its exceptional capability in efficiently sampling material configuration space, as shown by the generation of high-fidelity crystal structures that closely approximate their respective local minima, with average root mean squared displacements (RMSDs) between atom positions well below 1 \AA. The model achieves a high success rate and efficiency in predicting experimental structures without the need for density-functional-theory (DFT) based local optimization, and the results are comparable or superior to those from conventional CSP methods based on global optimization.

\begin{figure*}
\includegraphics[width=0.88\linewidth]{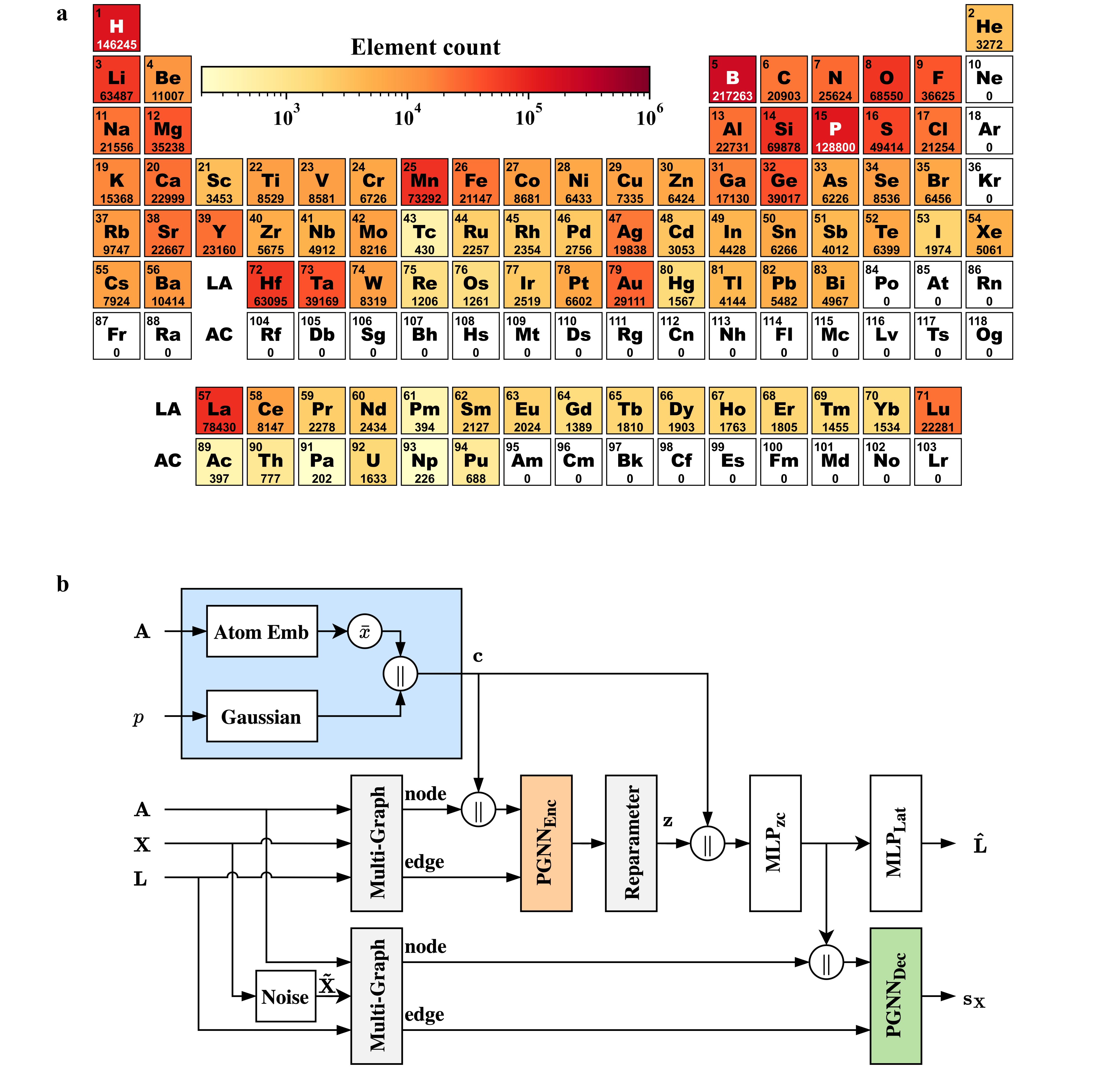}
\caption{\label{fig:fig1}
\textbf{Dataset and model architecture.} \textbf{a} Distribution of elements in the MP60-CALYPSO dataset. Color intensity represents the number of structures containing each element. \textbf{b} Model architecture of Cond-CDVAE. $\bar{x}$ denotes weighted average of element embedding vectors based on the number of atoms of each type, $\parallel$ denotes concatenation, and $\text{MLP}_{\square}$ denotes fully connected layers.}
\end{figure*}

\section*{Results}
\textbf{The MP60-CALYPSO dataset.} Crystal structures from MP~\cite{Jain.MaterialsProject.2013} (up to 60 atoms in a unit cell) and CALYPSO datasets are selected and combined to form the MP60-CALYPSO dataset, which is the basis for training and validating our GM (see Method for details on dataset curation). The MP dataset encompasses nearly all experimentally stable structures sourced from the Inorganic Crystal Structure Database (ICSD)~\cite{Zagorac.ICSD.2019}, supplemented by a diverse range of theoretical structures. The CALYPSO dataset comprises structures derived from prior CSP efforts, gathered from the wide CALYPSO community. The structures from both datasets have previously undergone DFT structural relaxations, ensuring that they are locally or globally stable. While all the structures from MP dataset were obtained at ambient pressure, those from the CALYPSO dataset are generally associated with high-pressure conditions. Integrating structures from both datasets enables a broad representation of chemical diversity inherent in inorganic materials across a wide range of pressures. We adopt relatively lenient criteria for selecting structures from the two datasets in an effort to collect reasonable bonding environments that are as diverse as possible (see Method).

\subfile{table1}

Fig.~\ref{fig:fig1}a and Table~\ref{tab:tab1} show the statistics of the MP60-CALYPSO dataset, which contains 670,979 crystal structures, spanning 86 elements, 85,824  chemical compositions, and 114,733 distinct structural prototypes. The MP dataset predominantly enriches this collection with its vast array of chemical compositions, whereas the CALYPSO dataset contributes a substantial diversity of structural prototypes. Notably, the dataset includes a considerable number of hydrogen- and boron-containing structures, with 146,245 and 217,263 instances, respectively, reflecting the research focus of the CALYPSO community, which has been directed toward superhydride superconductors and superhard materials.

\textbf{The Cond-CDVAE model.} Following the notation of Ref.~\cite{Xie.CDVAE.2022}, a crystal structure that contains $N$ atoms in a unit cell $\mathbf{M}= (\mathbf{A},\mathbf{X},\mathbf{L})$ is defined by the atom types $\mathbf{A} = (a_1, \dots ,a_N) \in \mathbb{A}^N$, atom coordinates {$\mathbf{X}=(\mathbf{x}_1,\dots,\mathbf{x}_N) \in \mathbb{R}^{N\times3}$}, and periodic lattice $\mathbf{L} = (\mathbf{l}_1,\mathbf{l}_2,\mathbf{l}_3) \in \mathbb{R}^{3 \times 3}$. Our objective is to train a GM that is conditioned on both chemical composition ($\mathbf{A}$) and pressure ($p$), drawing from the empirical distribution represented in the MP60-CALYPSO dataset $q(\mathbf{M}|\mathbf{A},p)$. To achieve this goal, we modified the CDVAE — a model that stands at the forefront of GM technology for crystal structures and operates within a VAE framework — to accommodate conditional generation. The architecture and workflow of the Cond-CDVAE are summarised in Fig.~\ref{fig:fig1}b. A multi-graph approach was used to represent the crystal structure. Both encoder and decoder were parameterized by SE(3) equivariant graph neural networks~\cite{Gasteiger.DimeNet.2020, Gasteiger.DimeNet++.2020, Gasteiger.GemNet.2021} to ensure permutation, translation, rotation, and periodic invariance. Additionally, the decoder was particularly designed as a noise conditional score network~\cite{Song.NCSN.2019}.

\subfile{table2}

During the training process, a periodic graph neural network encoder $\text{PGNN}_\text{Enc}(\mathbf{M})$ encodes $\mathbf{M}$ into a latent representation $\mathbf{z}$. A conditional module integrates discrete (chemical composition, $\mathbf{A}$) and continuous (pressure, $p$) attributes into a conditional vector $\mathbf{c}$.  A lattice predictor $\text{MLP}_\text{Lat}(\mathbf{z},\mathbf{c})$ takes the latent representation $(\mathbf{z},\mathbf{c})$ as input and predicts $\mathbf{\hat{L}}$, while a periodic graph neural network decoder $\text{PGNN}_\text{Dec}(\mathbf{\tilde{M}}|\mathbf{z},\mathbf{c})$ inputs a noisy structure $\mathbf{\tilde{M}}=(\mathbf{A},\mathbf{\tilde{X}},\mathbf{L})$ with coordinate noises $\mathbf{\sigma}_{\mathbf{X}}$, along with $(\mathbf{z},\mathbf{c})$, to output a score $\mathbf{s}_{\mathbf{X}} = \mathbf{s}_{\mathbf{X}}(\mathbf{\tilde{M}} | \mathbf{z},\mathbf{c};\sigma_{\mathbf{X}}) \in \mathbb{R}^{N \times 3}$, which denoise coordinates $\mathbf{\tilde{X}}$ to their ground truth values $\mathbf{X}$. The $\sigma_{\mathbf{X}}$ is sampled from a geometric sequence $\{\sigma_{\mathbf{X},j}\}_{j=1}^L$ of length $L$. The total loss of the model $\mathcal{L}_{\text{tot}}$ is a combination of a Kullback–Leibler (KL) divergence loss for the VAE $\mathcal{L}_{\text{KL}}$, a lattice loss $\mathcal{L}_{\text{Lat}}$, and a decoder denoising loss $\mathcal{L}_{\text{Dec}}$, written as follows,

\begin{alignat*}{1}
&\mathcal{L}_{\text{tot}} =
  \beta\mathcal{L}_{\text{KL}}
+ \lambda_{\mathbf{L}}\mathcal{L}_{\text{Lat}}
+ \lambda_{\mathbf{X}}\mathcal{L}_{\text{Dec}} \\
&\mathcal{L}_{\text{KL}} = -D_{KL}[q(\mathbf{z}|\mathbf{M},\mathbf{c})\Vert p(\mathbf{z}|\mathbf{c})] \\
&\mathcal{L}_{\text{Lat}} = \text{MSE}(\hat{\mathbf{L}},\mathbf{L}) \\
&\mathcal{L}_{\text{Dec}} = \frac{1}{2L}\sum\limits_{j=1}^{L} \\
&\quad \  [
\mathbb{E}_{q(\mathbf{M}|\mathbf{A},p)}\mathbb{E}_{q_{\sigma_{\mathbf{X},j}}(\mathbf{\tilde{M}}|\mathbf{M})}
(\Vert \mathbf{s_X} - \frac{\mathbf{d}_{\text{min}}(\mathbf{X},\mathbf{\tilde{X}})}{\sigma_{\mathbf{X},j}}\Vert_2^2)
] 
\end{alignat*}

\noindent where $\lambda_\mathbf{L}$, $\lambda_{\mathbf{X}}$, and $\beta$ are the respective weights assigned to each loss component, $\text{MSE}(\cdot,\cdot)$ represents the mean squared error, and $\mathbf{d}_{\text{min}}(\cdot,\cdot)$ denotes the minimum difference between coordinates considering periodicity.

In the generation process, the model integrates input chemical composition $\mathbf{A}$ and pressure $p$ into a conditional vector $\mathbf{c}$ and samples a $\mathbf{z}$ from the latent space. Using this sampled $\mathbf{z}$ along with the conditional vector $\mathbf{c}$, the lattice predictor $\text{MLP}_\text{Lat}(\mathbf{z},\mathbf{c})$ estimates the lattice vectors $\mathbf{L}$. This prediction is used to initialize a noisy crystal structure $\mathbf{\tilde{M}}=(\mathbf{A},\mathbf{\tilde{X}},\mathbf{L})$.  Subsequently, the decoder applies Langevin dynamics, conditioned on the $(\mathbf{z},\mathbf{c})$, to refine the noisy atomic positions $\mathbf{\tilde{X}}$ and produce the final crystal structure. Details for model implementation and training are provided in Methods.

\begin{figure*}
\includegraphics[width=1.0\linewidth]{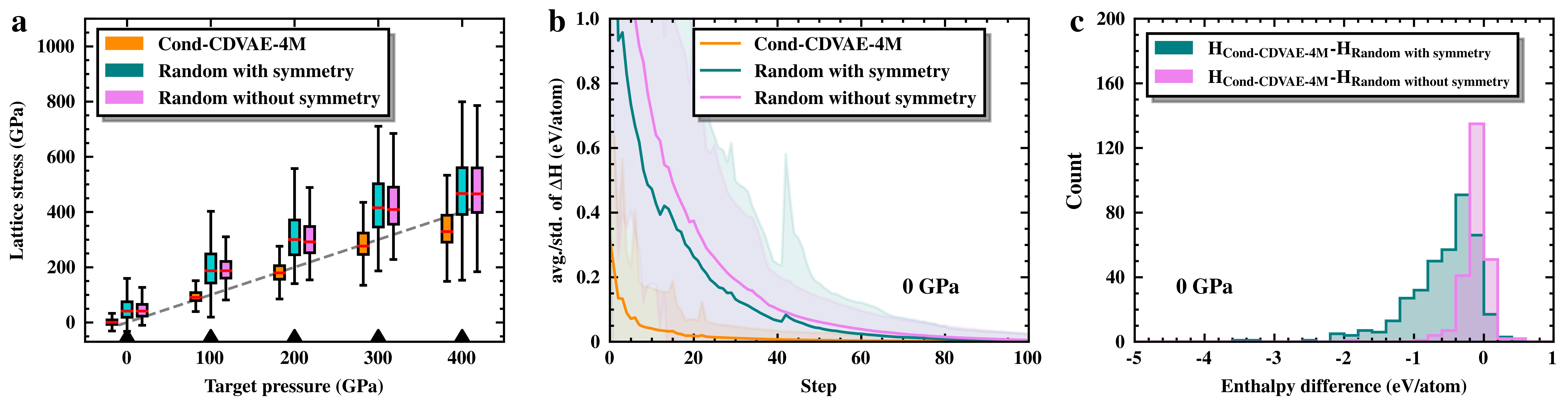}
\caption{\label{fig:fig2} \textbf{Comparison of random and model-generated structures.} \textbf{a} Boxplot illustrating lattice stress distributions at varying target pressures, with medians indicated by red lines, boxes encompassing the interquartile range, and whiskers reaching 1.5 times this range. \textbf{b} Average energy curves for structures versus ionic steps in DFT local optimizations at zero pressure, relative to the local minimum enthalpy, with shaded areas denoting standard deviation. \textbf{c} Energy distributions of optimized model-generated structures relative to those of randomly generated structures.}
\end{figure*}

\textbf{Reconstruction performance.} 
Within the framework of VAE, we initially evaluated the Cond-CDVAE's ability to reconstruct a crystal structure from its encoded latent representation. We trained the Cond-CDVAE on the MP60-CALYPSO dataset with two configurations to access the impact of model capacities on performance, one with 4 million parameters (referred to as Cond-CDVAE-4M) and another with 86 million parameters (referred to as Cond-CDVAE-86M); details on model training can be found in Methods. To compare with the baseline model CDVAE, we also train the Cond-CDVAE on the MP20 dataset in Ref~\cite{Xie.CDVAE.2022}, which consists of most experimental structures in the Material Project with up to 20 atoms in a unit cell. By contrasting the generated structures against the original inputs in the test sets, we calculated the match rate, defined as the proportion of structures successfully reconstructed, and the average normalized root mean square distance (N-RMSD) between matched structures, with results displayed in Table~\ref{tab:tab2} (see Methods for details on structure matching). For MP20, the Cond-CDAVE has a higher match rate than the CDVAE, which can be attributed to the fact that in Cond-CDVAE the composition of a structure is predetermined and does not participate in the diffusion process. Transitioning from the MP20 to the MP60-CALYPSO dataset results in a decline in the model's reconstruction performance, evidenced by a reduced match rate and an elevated N-RMSD. This suggests a heightened complexity within the MP60-CALYPSO dataset. However, the model's reconstruction performance can be improved by enhancing its capacity, indicating that additional training parameters can help the model better capture the intricacies of more complex datasets like MP60-CALYPSO. Moreover, we found that transitioning from the MP20 to the MP60-CALYPSO dataset slightly reduced the model's reconstruction performance on MP20 structures. For instance, the Cond-CDVAE-86M model trained on MP60-CALYPSO exhibits a match rate of 50.46\% on MP structures with fewer than 20 atoms per unit cell in its test set. This match rate is only marginally lower than that of the model trained exclusively on the MP20 dataset, which has a match rate of 57.75\%.

\subfile{table3}

\textbf{Generation performance.} We then proceed to evaluate the model's generation performance, which is more important than reconstruction performance for practical CSP applications. In the following, we concentrate exclusively on the results of the Cond-CDVAE models trained on MP60-CALYPSO, as these are capable of generating structures under specified pressures. This capability is crucial for simulating realistic conditions that materials would encounter in practical settings, thereby making the Cond-CDVAE model particularly relevant for CSP tasks. 

We generate 500 structures using Cond-CDVAE-4M at pressures of 0, 100, 200, 300, and 400 GPa. The chemical compositions for these structures are randomly selected from the test set to ensure a broad representation. For comparison, alongside each model-generated structure, we also produce a structure both with and without constraints of space-group symmetries at the same composition and cell volume using the CALYPSO structure prediction package~\cite{Wang.CALYPSO.2012, Shao.Symmetry-CALYPSO.JCP.2022}. All these structures are subjected to DFT single-point and local optimization calculations to assess their quality. The distribution of lattice stresses for both model-generated and randomly produced structures at each pressure is depicted in Fig.~\ref{fig:fig2}a. The median lattice stresses of the model-generated structures show good alignment with the target pressures, indicating that the model effectively learns the impact of pressure on the structures. The variance increases with pressure, correlating with the reduced volume of training data at higher pressures. Moreover, as a material's incompressibility increases with pressure, the same magnitude of error in predicted lattice parameters would result in a larger discrepancy in stress at higher pressures. Notably, even if the same cell volume is used as in each model-generated structure, the distribution for randomly produced structures is broader, and their median lattice stresses are not as closely aligned. This improved distribution for model-generated structures is attributed to a more reasonable prediction of lattice parameters and atomic positions. 

The quality of generated structures can be further evaluated by their performance in DFT local optimization, as indicated by the convergence rate and the number of ionic steps. The convergence rate represents the percentage of structures that have been successfully optimized by VASP code (see Method), while an ionic step refers to an iteration during which the positions of atoms are adjusted to minimize the system's total energy. An unreasonable structure typically causes a calculation failure or requires a large number of ionic steps to converge. Therefore, these two metrics effectively reflect the quality of structures, which is particularly useful in CSP scenarios where numerous structures need to be optimized. As shown in Table~\ref{tab:tab3}, the superior quality of model-generated structures is evidenced by significantly higher convergence rates compared to their randomly produced counterparts, both with or without symmetries. The average RMSDs between model-generated structures and their local minima across different pressures range from 0.48 to 0.79 \AA, which are significantly lower than those of randomly produced structures and comparable to those obtained by the original CDVAE in Ref.~\cite{Zeni.MatterGen.2023} (see Methods for details on structure matching and RMSD calculations). This result confirms that the model-generated structures are more accurately approximated and closer to their respective local energy minima. The average number of ionic steps required to reach convergence for model-generated structures is comparable to that of randomly produced structures that incorporate symmetries and significantly fewer than for non-symmetric randomly produced structures. The reduced number of ionic steps needed for symmetric randomly produced structures can be attributed to the fewer degrees of freedom that need to be optimized due to their inherent symmetry. In contrast, the efficiency of model-generated structures is attributed to their initial configurations, which are already in closer proximity to the local energy minima, thus necessitating fewer adjustments to achieve convergence. These observations are further illustrated by the average energy curves during local optimization, as depicted in Fig.~\ref{fig:fig2}b for 0 GPa and Supplementary Fig.~S4 for other pressure conditions. Moreover, the enthalpy (reduced to potential energy at 0 GPa) distribution of the optimized model-generated structures, as shown in Fig.~\ref{fig:fig2}c, suggests that a majority of the relaxed model-generated structures possess superior thermodynamic stability relative to their randomly produced counterparts.

\begin{figure}
\includegraphics[width=0.80\linewidth]{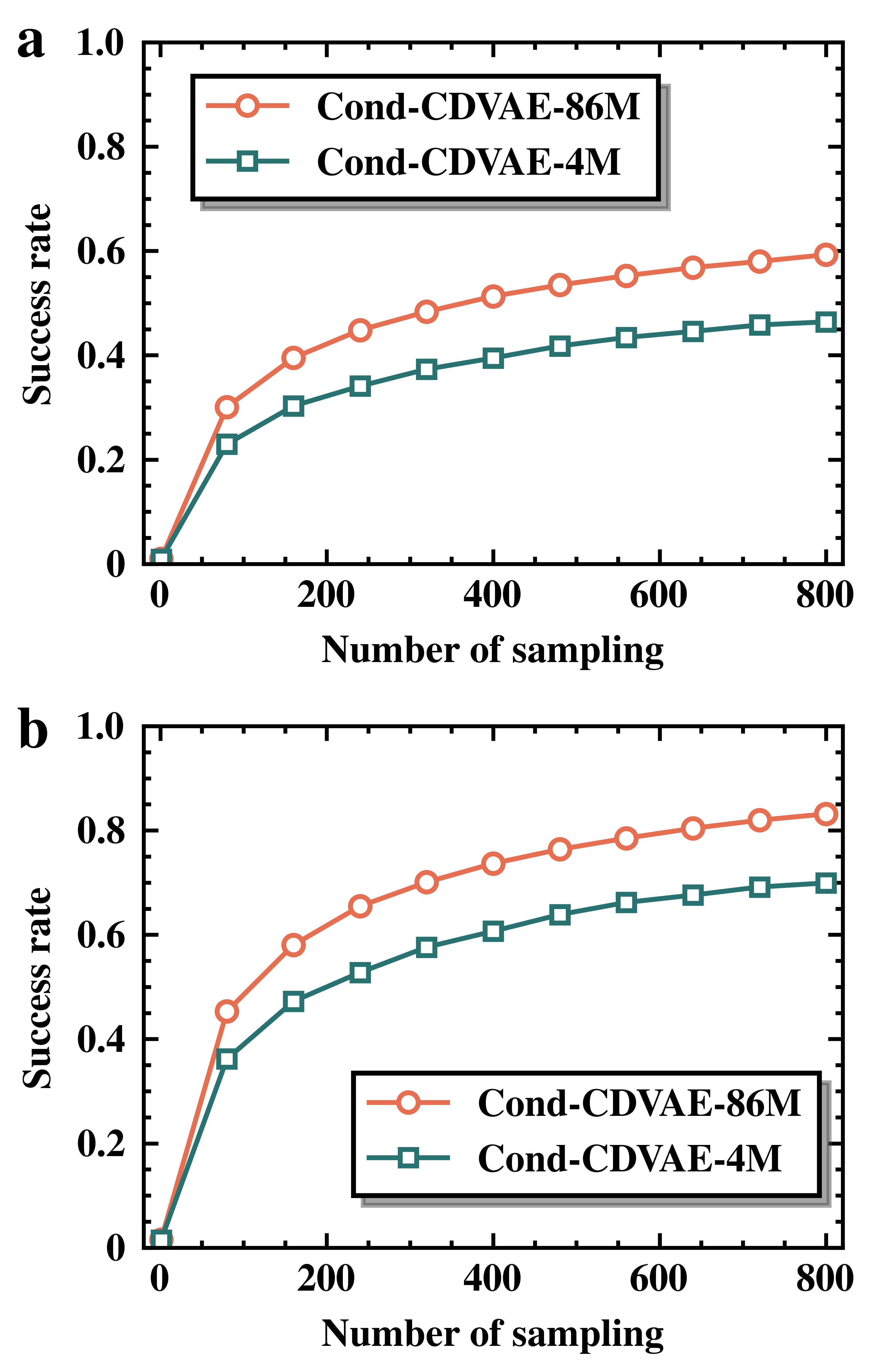}% Here is how to import EPS art
\caption{\label{fig:fig3} \textbf{The success rate of the Cond-CDVAE model on structure prediction of experimental structures as a function of number of structural samplings.} Results for \textbf{a} all 3,547 MP experimental structures in the test set and \textbf{b} 2,062 MP experimental structures in the test set containing fewer than 20 atoms in a unit cell.}
\end{figure}

\subfile{table4}

\begin{figure*}
\includegraphics[width=0.95\linewidth]{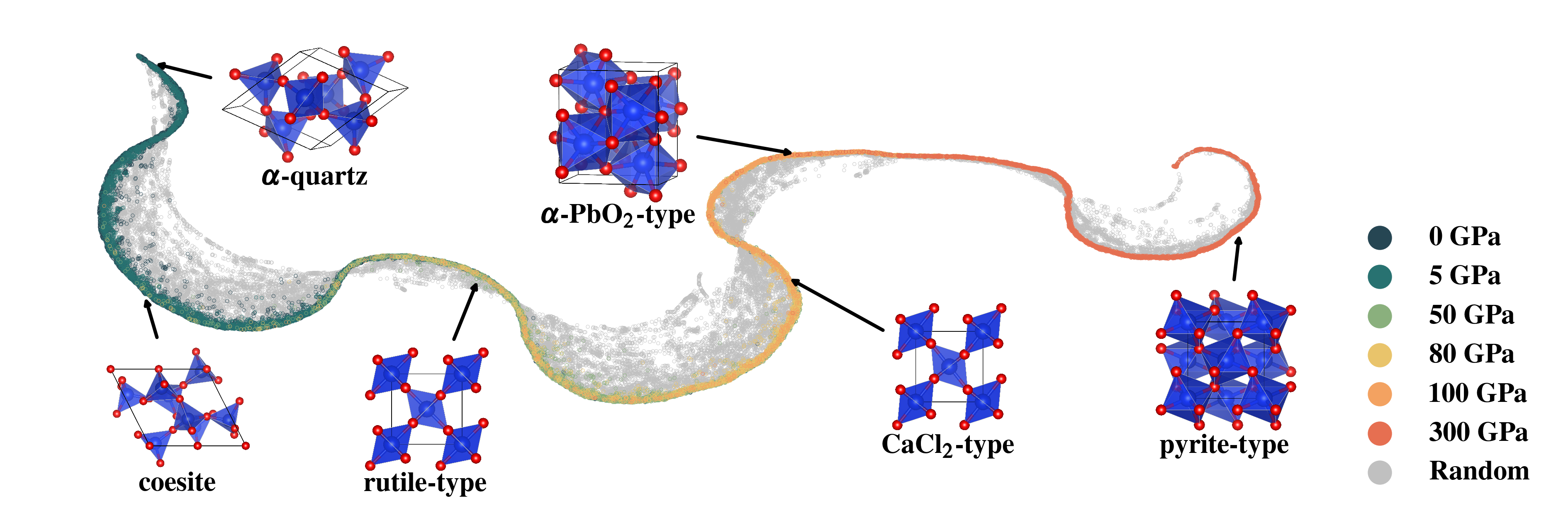}
\caption{\label{fig:fig4}\textbf{Two-dimensional projection of silica structures generated by the Cond-CAVAE-4M model alongside random symmetric structures.} Data points representing the model-generated structures are color-coded to reflect the input pressure conditions, whereas the data points for the random structures are shown in grey. Ground-state structures of silica at various pressures are also depicted.}
\end{figure*}

\textbf{Structure prediction performance.} Finally, we assess the performance of Cond-CDVAE in practical CSP tasks. We begin with the model's performance on 3,547 ambient-pressure experimental structures, which are MP structures with an ICSD label in our test set. For each chemical system, we use Cond-CDVAE to generate up to 800 structures and monitor whether and when the ground truth is successfully reproduced. A structure prediction is considered unsuccessful if the ground truth is not reproduced within these samplings. A slightly stricter criterion for structure matching was adopted compared to that used for assessing reconstruction performance (see Methods). The success rate, defined as the percentage of systems successfully reproduced within a certain number of samplings, is shown in Fig.~\ref{fig:fig3}a. With Cond-CDVAE-4M, we observe that the success rate incrementally rises as the number of samplings increases, plateauing at around 46.4\%. Meanwhile, Cond-CDVAE-86M has a higher success rate of 59.3\%, suggesting that increased model capacity would enhance the model's CSP ability. Further examination indicates that systems with a larger number of atoms are more challenging to predict.  As illustrated in Fig.~\ref{fig:fig3}b, the success rate for MP experimental structures containing fewer than 20 atoms (2,062 systems) achieves high values of 69.9\% and 83.2\% for Cond-CDVAE-4M and Cond-CDVAE-86M, respectively. Detailed analysis of the CSP results, as shown in Supplementary Fig. S5, also substantiates this observation, demonstrating that the success rate is closely related to the number of atoms in the test system, decreasing monotonically with the number of atoms. This is partially attributable to the limited volume of training data available for systems with a larger number of atoms (see Supplementary Fig.~S3 for statistics on the MP60-CALYPSO dataset). The lower success rate for large complex systems implies the classic challenge of CSP as the complexity of corresponding PES increases exponentially with system size. Nevertheless, the Cond-CDVAE model demonstrates commendable CSP capabilities for materials at ambient pressure. It should be noted that the approach and parameters used for structure matching are relatively lenient, as they are consistent with previous work to facilitate comparison \cite{Xie.CDVAE.2022,Jiao.DiffCSP.2023}. When space group determination is included in the structure matching process, the success rate slightly decreases, as shown in Supplementary Fig.~S6. In theory and common practice, two structures could be considered identical if they can be locally optimized to the same local minimum through ab initio calculations. However, this approach is computationally prohibitive for the large-scale statistical analysis in this work.

We then broaden our evaluation of the model's CSP performance by focusing on three representative systems — lithium (Li), boron (B), and silica (\ce{SiO2}) — all known for displaying complex polymorphic behaviors under high-pressure conditions. For lithium (Li), we select the cubic $cI16$ phase, which becomes stable at pressures above approximately 40 GPa, exemplifying the complex high-pressure phases of alkali metals~\cite{Hanfland.2000}. For B, we focus on three prominent phases: $\alpha$-\ce{B12}, $\gamma$-\ce{B28}, and $\alpha$-Ga-type B, stable at ambient conditions, above 19 GPa, and 89 GPa, respectively~\cite{Oganov.gammaB28.2009}. Both $\alpha$-\ce{B12} and $\gamma$-\ce{B28} contain the characteristic \ce{B12} icosahedra, whereas in $\alpha$-Ga-type B, the icosahedra are disrupted to achieve denser packing. For silica, we select six stable phases that cover pressures up to 300 GPa: the $\alpha$-quartz and coesite phases stable below 10 GPa, characterized by \ce{SiO4} tetrahedra, and rutile-type, \ce{CaCl2}-type, $\alpha$-\ce{PbO2}-type, and pyrite-type phases stable at higher pressures featuring a sixfold coordination number of Si~\cite{Liu.silica.PRL.2021}. All these phases are manually removed from the training set.

For each test system, we conduct structure predictions utilizing both the Cond-CAVAE-4M model and the CALYPSO code. Multiple independent simulations are carried out, consisting of five runs for the Cond-CAVAE-4M model and three for the CALYPSO code (See Methods for details). A run is deemed successful if it reproduces the ground state in 1,000 samplings. The summarized test results in Table~\ref{tab:tab4} highlight the CSP capabilities of the Cond-CAVAE model. Specifically, for all the silica test systems, the Cond-CAVAE-4M model consistently required fewer samplings to locate the global minimum compared to CALYPSO. Notably, for the complex coesite phase and the $\gamma$-\ce{B28}, which comprise 24 and 28 atoms per unit cell, respectively, the model successfully reproduces the ground states with an average of 328.0 and 341.0 samplings, respectively, while the CALYPSO does not succeed within 1,000 samplings. However, the Cond-CAVAE-4M model shows limitations in efficiently locating the $cI16$ Li and $\alpha$-\ce{B12} structures, requiring more samplings or yielding lower success rates. This highlights areas for potential enhancement of the model. For example, $cI16$ Li is characterized by a high space-group symmetry of $I\bar{4}3d$, with 16 atoms within each unit cell situated at the $16c$ $(x,x,x)$ Wyckoff positions. CALYPSO's superior performance in identifying this structure can be primarily attributed to its symmetry considerations during the prediction process, which significantly reduces the system's degrees of freedom. Incorporating space-group symmetry considerations into the model could, therefore, be a valuable strategy to augment its CSP performance. It should be noted that the small number of tests for each system may not be sufficient to achieve statistically significant results. For example, a lower sampling number for $\alpha$-\ce{B12} (74) for the model is based solely on one successful run. The difficulty for the model to predict this structure is reflected in the success rate of 1/5.

To further evaluate the characteristics of structures generated by the model, we conduct dimensionality reduction on the silica structures through manifold learning, utilizing the SHEAP~\cite{Shires.SHEAP.2021} method. For comparison, we also create a randomly produced structure with identical volume and symmetry constraints using CALYPSO. The projection of these structures onto a two-dimensional plane is illustrated in Fig.~\ref{fig:fig4}. The distinctive attributes of the model-generated structures are apparent. The data points for these structures are closely grouped and display an orderly pattern that aligns with varying pressures. In contrast, the data points corresponding to the randomly generated structures are more scattered, suggesting that the model has effectively captured the distribution of local energy minima in the dataset, which exhibits a low-dimensional feature. This finding highlights the model's proficiency in distilling complex distributions into a more manageable, lower-dimensional space.

\section*{Discussion}
Leveraging the latest advances in crystal structure generation through GMs and an extensive collection of crystal structure data, we devised a universal GM capable of producing valid and varied crystal structures tailored to allow user-specified physical conditions such as composition and pressure. The resulting Cond-CDVAE model demonstrates excellent proficiency in creating physically meaningful crystal structures and exhibits superior CSP performance compared to state-of-the-art methods based on global optimization algorithms under both ambient and high-pressure conditions. Despite these achievements, there is significant potential for further enhancements in the model and its underlying database. As demonstrated by the CSP tests on the $cI16$ Li and $\alpha$-\ce{B12} structures, integrating space-group symmetry, a fundamental aspect of crystal structures, into GMs is highly desirable to substantially improve the quality of generated structures, thereby enhancing the GM's performance in CSP tasks. Moreover, structures from the CALYPSO dataset, collected from the wide user community and computed under different DFT codes and settings, with more than half not considered in this work based on a simple filter criterion, suggest that a more thorough refinement and filtering of this dataset could further improve the model's performance. The conditional generation strategy employed here is not limited to composition and pressure; it could be extended to other material properties, such as mechanical and electronic properties, as well as thermodynamic stability, provided there are suitable results available across the crystal structure database. We are committed to ongoing efforts to refine both the model and database. The methodologies and benchmarks introduced here are poised to serve as invaluable references for future work in crystal structure generation within the realm of CSP.

The development of GM in materials science is rapidly evolving. Recent progresses, particularly those employing diffusion models, have facilitated integration of both lattice and atomic positions into the structure generation process~\cite{Zeni.MatterGen.2023, Jiao.DiffCSP++.2024}, which has greatly improved the quality of the generated structures. Moreover, the integration of space-group symmetry into GMs is also on the horizon~\cite{Zhao.PGCGM.2023, Jiao.DiffCSP++.2024}. With continuous progress in model architectures and improvements in the quantity and quality of crystal databases, GMs, combined with the latest machine learning potentials~\cite{Xie.CGCNN.2018, Chen.M3GNet.2022, Batzner.NequIP.2022, Deng.CHGNet.2023, Zhang.DPA-2.2023, Batatia.MACE-MP-0.2023, Merchant.GNoME.2023}, are expected to become an indispensable component of modern CSP toolkits.

\section*{Methods}

\textbf{Dataset Curation.} The raw data were sourced from the MP dataset as of February 13, 2023, and from the CALYPSO dataset, which was compiled from previous structure search calculations contributed by the CALYPSO community, containing 154,718 and 2,804,206 structures, respectively. To enable the GMs to generate the widest possible variety of crystal structures — a critical requirement in structure search calculations for a thorough exploration of the energy landscape — we applied relatively lenient criteria when selecting structures from both datasets. For the MP dataset, after removing duplicates, we first filter out erroneous structures based on interatomic distance and cell volume. Specifically, we considered a structure implausible if the shortest interatomic distance fell below 0.5 or exceeded 5.2 \AA, considering that the maximum covalent radius among the elements is approximately 2.6 \AA. Additionally, any structure that had a cell volume exceeding four times that of the theoretical volume of a hard sphere model — calculated using the covalent radii — was also excluded from our dataset. Then we selected structures with up to 60 atoms in a unit cell, a negative formation energy, and an energy above hull no greater than 0.2 eV/atom. Following these criteria, a total of 99,243 structures were selected from the MP dataset. For the CALYPSO dataset, we applied similar criteria to remove duplicates and erroneous structures. However, we modified the cell volume threshold to no more than double the theoretical volume predicted by a hard sphere model, taking into account that many of these structures are typically relaxed at elevated pressures. We then selected half of the structures with the lowest energies from each structure search calculation. Consequently, a total of 571,736 structures are selected from the CALYPSO dataset. Structural prototypes were determined by \texttt{\detokenize{XtalFinder}} in AFLOW~\cite{Hicks.AFLOW.2021}. We employed a random data split, allocating 80\% of the data for training, 10\% for validation, and the remaining 10\% for testing. 

\textbf{Structure matching.} We utilized the \texttt{\detokenize{StructureMatcher}} function in the pymatgen~\cite{Ong.pymatgen.2013} library to match structures and measure their similarity. For detecting duplicate structures in dataset curation and assessing the structure prediction performance, we considered two structures identical if they can be matched using the default parameters: \texttt{\detokenize{ltol=0.2}}, \texttt{\detokenize{stol=0.3}}, \texttt{\detokenize{angle_tol=5}}. In evaluating the model's reconstruction performance, we employed the same criteria as in the original CDVAE work: \texttt{\detokenize{ltol=0.3}}, \texttt{\detokenize{stol=0.5}}, \texttt{\detokenize{angle_tol=10}} for consistency \cite{Xie.CDVAE.2022}. For matched structures, the root mean squared displacements between matched atom pairs were calculated and normalized by $\sqrt[3]{\bar{V}/N}$, where $\bar{V}$ is the volume calculated using the average lattice parameters, denoted as N-RMSD. 

In assessing the generation performance, similarity between unrelaxed and relaxed structures is measured by the root mean squared displacements between atom pairs under minimum image convention of the periodic boundary condition without 3D superposition or volume normalization, denoted as RMSD.

We acknowledge the recent development of the CSPBenchMetrics package~\cite{Wei_CSPBenchMetrics_2024}, which integrates 12 metrics for measuring structural similarity in CSP. However, to facilitate the comparison of our model with the latest state-of-the-art models, we have exclusively used pymatgen \texttt{\detokenize{StructureMatcher}} for structure matching. We plan to explore the use of stricter metrics in future work.

\textbf{Model implementation and training.} The Cond-CDVAE model is an extension of the original CDVAE, integrating the sophisticated DimeNet++ \cite{Gasteiger.DimeNet.2020, Gasteiger.DimeNet++.2020} as the encoder and GemNet-dQ \cite{Gasteiger.GemNet.2021} as the decoder. This model was enhanced with a conditional network to support conditional generation. Within this network, each atom type $a \in \mathbb{A}$ is represented by a distinct element embedding vector derived from an embedding layer. The chemical composition's collective representation was then determined by weighted average of these element embedding vectors based on the number of atoms of each type. Moreover, a continuous scalar property such as pressure ($p$) was standardized and expanded through a set of equidistant Gaussian bases, $\exp[-\frac{(p-\mu)^2}{2\sigma ^2}]$, where $\mu \in \{\mu_1, \mu_2, ... \}$ and $\sigma \in \{ \sigma_1, \sigma_2, ... \}$~\cite{Gebauer.cG-SchNet.2022}. Representations of chemical composition and pressure were then concatenated to form the conditional vector $\mathbf{c}$, which was passed into both the encoder and decoder. In practical implementation, we found that adding a tanh activation layer at the end of the encoder significantly improves the training stability compared with the original CDVAE.

Two models with different capacities were trained to investigate the effect of model size on performance. The first consists of four message-passing layers with hidden dimensions set to 128, comprising 4 million training parameters, thus referred to as the 4M-model. The second is constructed with six message-passing layers, each with hidden dimensions of 512, and it encompasses a total of 86 million training parameters, referred to as the 86M-model. Hyperparameters $\lambda_{\mathbf{L}}$, $\lambda_{\mathbf{X}}$, and $\beta$ were set to 10, 10, and 0.01 respectively. All models were trained with a batch size of 128, and a \texttt{ReducedLROnPlateau} strategy for adjusting the learning rate from 1e-4 reducing to 1e-5 with a factor of 0.6 and patience 30.

\textbf{Randomly generated structure via CALYPSO.} To assess generation performance, randomly generated structures were produced using the CALYPSO code~\cite{Wang.CALYPSO.2012, Shao.Symmetry-CALYPSO.JCP.2022} with the same compositions and cell volumes for comparison, both with and without constraints of space-group symmetries. For the generation of symmetric structures, one of the 230 space groups was randomly selected. Subsequently, lattice parameters and atomic coordinates were randomly generated within the constraints of the chosen space group. The minimum interatomic distance between atom A and B was set according to $0.7\times(R_\mathrm{A}+R_\mathrm{B})$, where $R_\mathrm{A}$ and $R_\mathrm{B}$ are \texttt{\detokenize{RCORE}} in POTCAR files of each element, respectively.

The structure search calculations for high-pressure phases of Li, B, and SiO$_2$ were conducted using the swarm-intelligence-based CALYPSO method~\cite{Wang.CALYPSO.2010} and the Cond-CDVAE-4M model. For the runs in CALYPSO, the population size was set to 30, and 1,000 structures were sampled, with 40\% generated randomly with constraints of symmetry, and 60\% produced by the PSO algorithm. The simulation cells were based on the number of atoms in the primitive unit cell of the known ground states. The underlying DFT local optimizations were performed using the VASP code~\cite{Kresse.VASP.1996}. For runs using the Cond-CDVAE model, the simulation cells were based on the number of atoms in both the primitive and conventional unit cells, and the optimal results are adopted in Table~\ref{tab:tab4}.

\textbf{DFT calculations.} DFT calculations were performed with the Perdew-Burke-Ernzerhof (PBE) exchange-correlation functional~\cite{Perdew.PBE.1996} and all-electron projector-augmented wave method~\cite{Blochl.PAW.1994}, as implemented in the VASP code~\cite{Kresse.VASP.1996}. An energy cutoff of 520 eV and a Monkhorst-Pack k-point sampling grid spacing of 0.25 \AA$^{-1}$ were used to ensure the convergence of the total energy. The default settings of PBE functional, Hubbard U corrections, and ferromagnetic initializations in pymatgen \texttt{\detokenize{MPRelaxSet}} function were employed, except opting for the \texttt{\detokenize{W_sv}} potential for tungsten. The maximum optimization ionic step and the maximum running time are constrained to 150 steps and 20 hours, respectively.

\textbf{Manifold visualization.} The dimensionality reduction and manifold learning were performed by the SHEAP method~\cite{Shires.SHEAP.2021} with the use of SOAP descriptor~\cite{Bartok.SOAP.2013}. The projection used a perplexity of 10, with a similarity threshold of 0.1 and a default minimum hard-sphere radius of 0.01. DScribe~\cite{Himanen.dscribe.2019} package was utilized to calculate the SOAP descriptors with parameters $n_{\text{max}},l_{\text{max}},r_{\text{cut}},\sigma = 5,4,6,1$.

\section*{Data availability}

The authors declare that the main data supporting the findings of this study are contained within the paper and its associated Supplementary Information.

\section*{Code availability}

The Cond-CDVAE source code is available on GitHub (\href{https://github.com/ixsluo/cond-cdvae}{https://github.com/ixsluo/cond-cdvae}). The CALYPSO code is free for academic use, by registering at \href{http://www.calypso.cn}{http://www.calypso.cn}.

\section*{References}
% Produces the bibliography via BibTeX.
% \bibliographystyle{naturemag}  % apsrev4-2
% \bibliography{ref}

\begin{acknowledgments}
The work is supported by the National Key Research and Development Program of China (Grant No. 2022YFA1402304, 2022YFA1405500, and 2023YFA1406200), the National Natural Science Foundation of China (Grants No. 12034009, 12374005, 52090024, 52288102, T2225013, 12174142, 22131006, 12104176, and U22A2098), the Fundamental Research Funds for the Central Universities, the Program for JLU Science and Technology Innovative Research Team. We want to thank ``Changchun Computing Center" and ``Eco-Innovation Center" for providing inclusive computing power and technical support of MindSpore during the completion of this paper. Part of the calculation was performed in the high-performance computing center of Jilin University. We also want to thank all CALYPSO community contributors for providing the structural data to construct the dataset.
\end{acknowledgments}

\section*{Author contributions}
J.L., Y.W., C.C., and Y.M. designed the research. X.L. wrote the code and trained the model. Z.W. and X.L. collected the dataset. X.L., Z.W., and P.G. conducted the calculations. X.L., Z.W., P.G., J.L., Y.W., C.C., and Y.M. analyzed and interpreted the data, and contributed to the writing of the paper.

\section*{Competing interests}

The authors declare no competing interests.

\section*{Additional information}

\textbf{Supplementary information} The online version contains supplementary material available at XXX.

\textbf{Correspondence} and requests for materials should be addressed to Jian Lv, Yanchao Wang, Changfeng Chen, or Yanming Ma.

\end{document}

\title{Supplementary Information for \texorpdfstring{\\}{}
Deep learning generative model for crystal structure prediction}

\author{Xiaoshan Luo}
\affiliation{Key Laboratory of Material Simulation Methods and Software of Ministry of Education, College of Physics, Jilin University, Changchun 130012, P. R. China}
\affiliation{State Key Laboratory of Superhard Materials, College of Physics, Jilin University, Changchun 130012, P. R. China}

\author{Zhenyu Wang}
\affiliation{Key Laboratory of Material Simulation Methods and Software of Ministry of Education, College of Physics, Jilin University, Changchun 130012, P. R. China}
\affiliation{International Center of Future Science, Jilin University, Changchun, 130012, P. R. China}

\author{Pengyue Gao}
\affiliation{Key Laboratory of Material Simulation Methods and Software of Ministry of Education, College of Physics, Jilin University, Changchun 130012, P. R. China}

\author{Jian Lv}
\email{lvjian@jlu.edu.cn}
\affiliation{Key Laboratory of Material Simulation Methods and Software of Ministry of Education, College of Physics, Jilin University, Changchun 130012, P. R. China}

\author{Yanchao Wang}
\email{wyc@calypso.cn}
\affiliation{Key Laboratory of Material Simulation Methods and Software of Ministry of Education, College of Physics, Jilin University, Changchun 130012, P. R. China}

\author{Changfeng Chen}
\email{chen@physics.unlv.edu}
\affiliation{Department of Physics and Astronomy, University of Nevada, Las Vegas, NV 89154, USA}

\author{Yanming Ma}
\email{mym@jlu.edu.cn}
\affiliation{Key Laboratory of Material Simulation Methods and Software of Ministry of Education, College of Physics, Jilin University, Changchun 130012, P. R. China}
\affiliation{International Center of Future Science, Jilin University, Changchun, 130012, P. R. China}

\date{\today}

\maketitle

% \section*{\huge{Supplementary Information}}
% \section*{Deep learning generative model for crystal structure prediction}
% \section*{\mdseries{X. Luo et al.}}

% \clearpage
\section*{Supplementary Figures}
\begin{figure}[!h]
\includegraphics[width=1.0\linewidth]{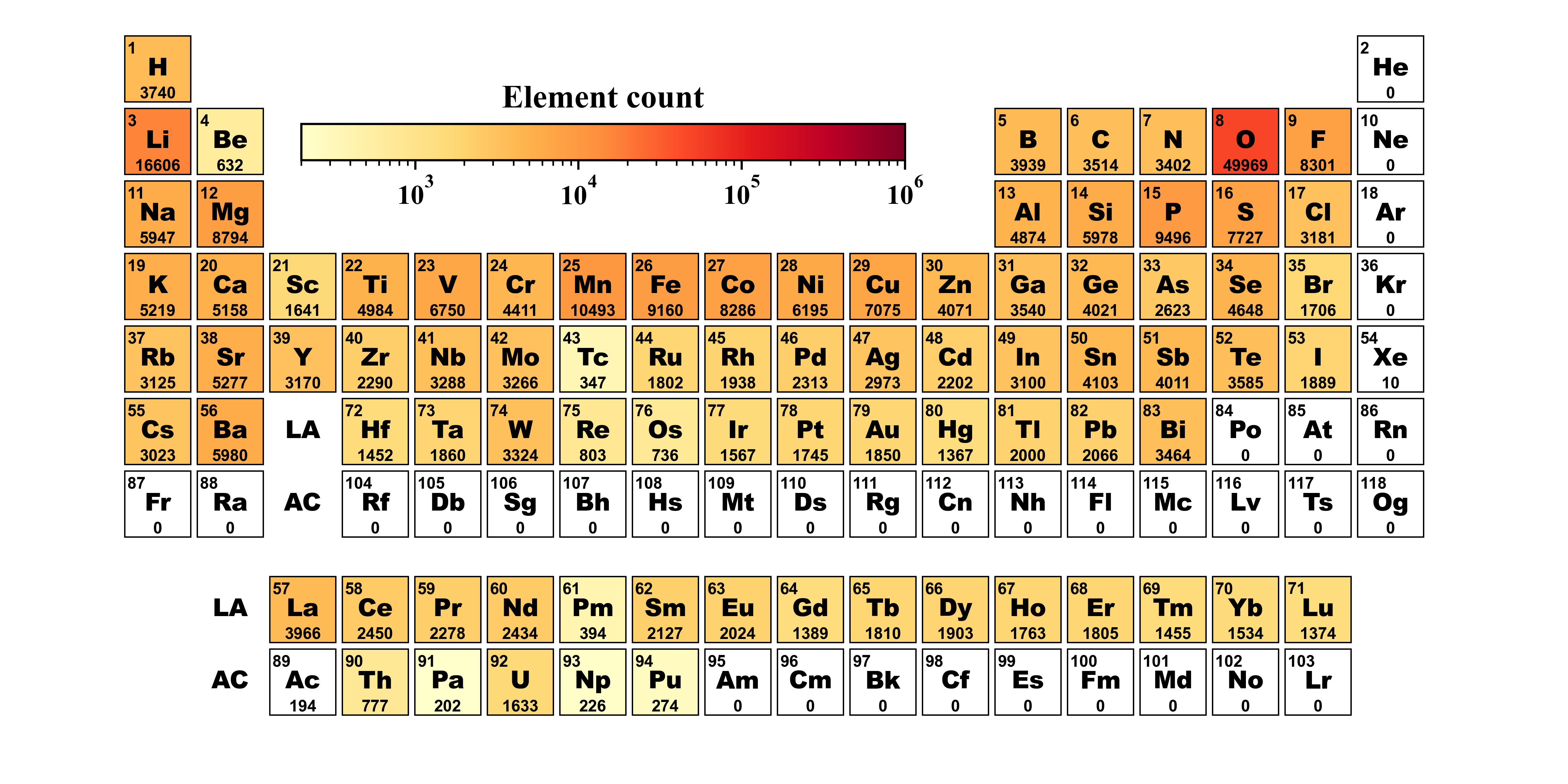}
\caption{\label{fig:sfig1}Distribution of elements for structures in MP60-CALYPSO contributing from Materials Project. Color intensity represents the number of structures containing each element.}
\end{figure}

\clearpage
\begin{figure}[!h]
\includegraphics[width=1.0\linewidth]{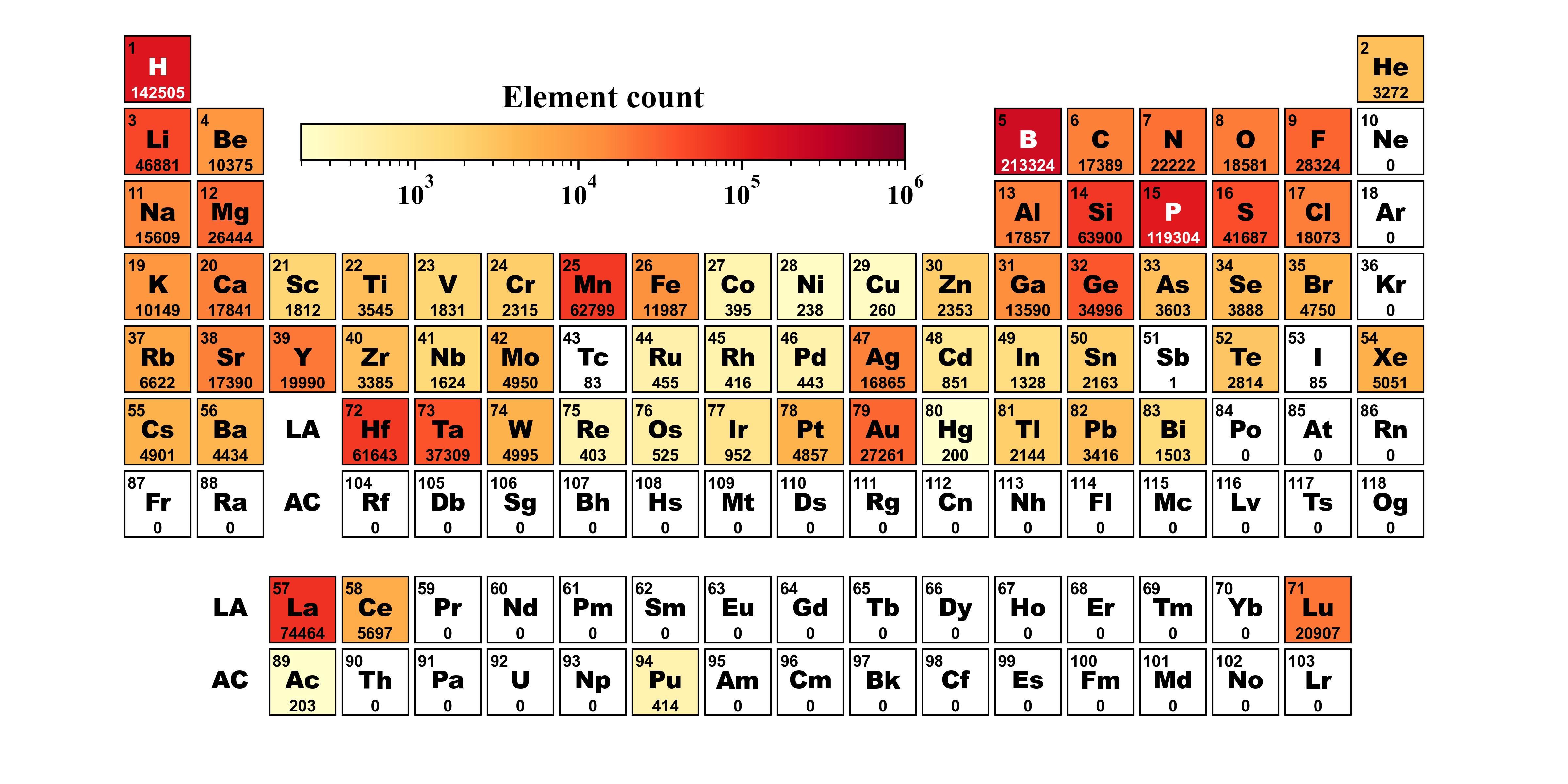}
\caption{\label{fig:sfig2}Distribution of elements for structures in MP60-CALYPSO contributing from CALYPSO community. Color intensity represents the number of structures containing each element.}
\end{figure}

\clearpage
\begin{figure}[!h]
\includegraphics[width=0.6\linewidth]{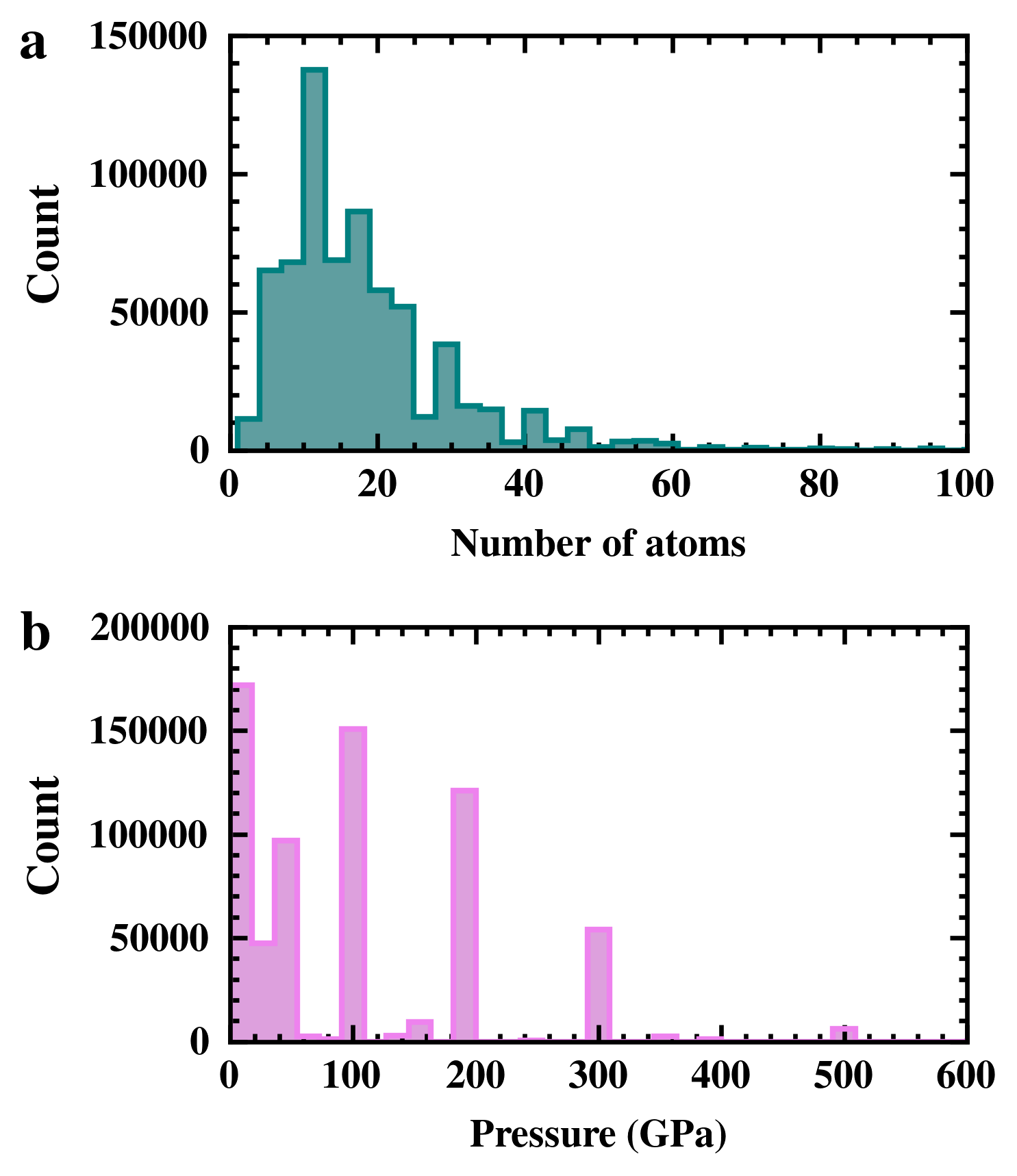}
\caption{\label{fig:sfig3}Distribution of \textbf{a} the number of atoms in a unit cell, and \textbf{b} the equilibrium pressure in the MP60-CALYPSO dataset}
\end{figure}

\clearpage
\begin{figure}[!ht]
\includegraphics[width=0.85\linewidth]{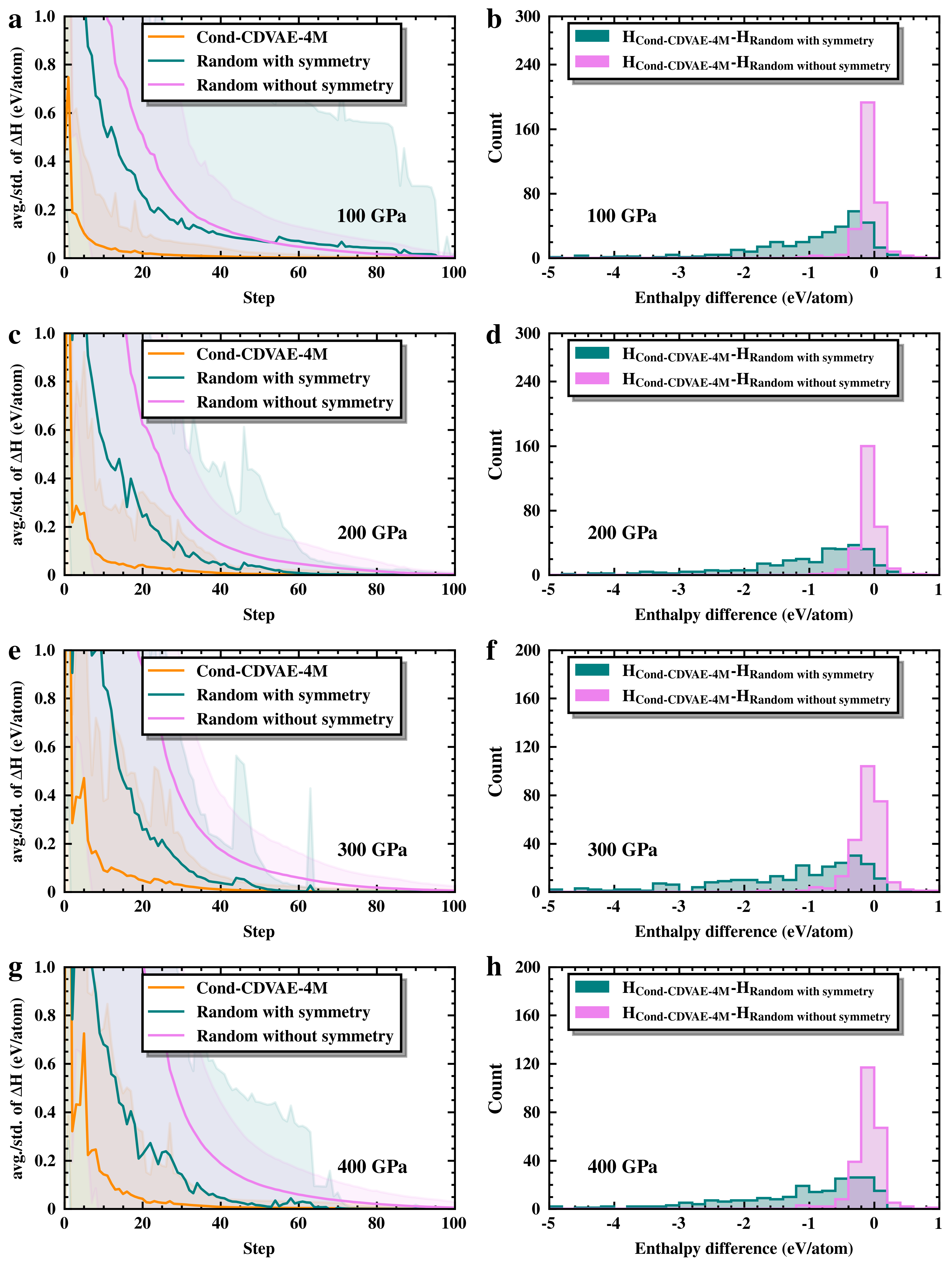}
\caption{\label{fig:sfig4}Comparison of random and model-generated structures. Displayed from top to bottom are structures generated at pressures of 100, 200, 300, and 400 GPa. \textbf{a,c,e,g} Average energy curves for structures versus ionic steps in DFT local optimizations relative to their local minimum enthalpy, with shaded areas denoting standard deviation. To mitigate the impact of erroneous ion steps exhibiting substantial energy fluctuations, steps with energy variations exceeding 10 eV/atom are ignored. \textbf{b,d,f,h} Energy distributions of optimized model-generated structures relative to those of randomly generated structures.}
\end{figure}

\clearpage
\begin{figure}[!ht]
\includegraphics[width=0.55\linewidth]{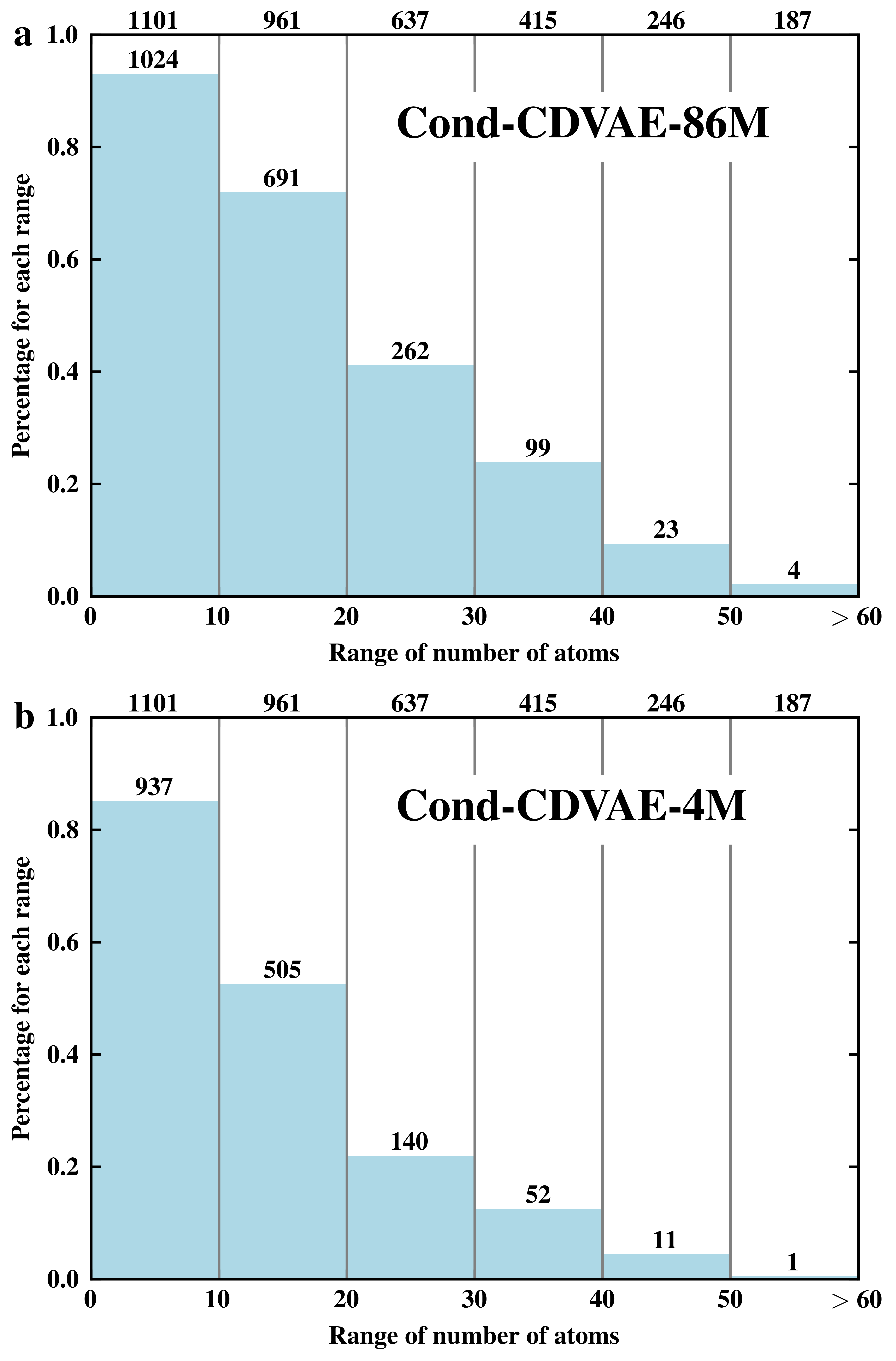}
\caption{\label{fig:sfig5} Percentages of successfully predicted ambient-pressure experimental structures in the test set by \textbf{a} Cond-CDVAE-86M and \textbf{b} Cond-CDVAE-4M, categorized by the range of the number of atoms in a unit cell. The numbers atop each bar indicate the number of structures successfully reproduced, while the numbers at the top of each bin represent the total number of structures within that range.}
\end{figure}

\clearpage
\begin{figure}[!ht]
\includegraphics[width=0.7\linewidth]{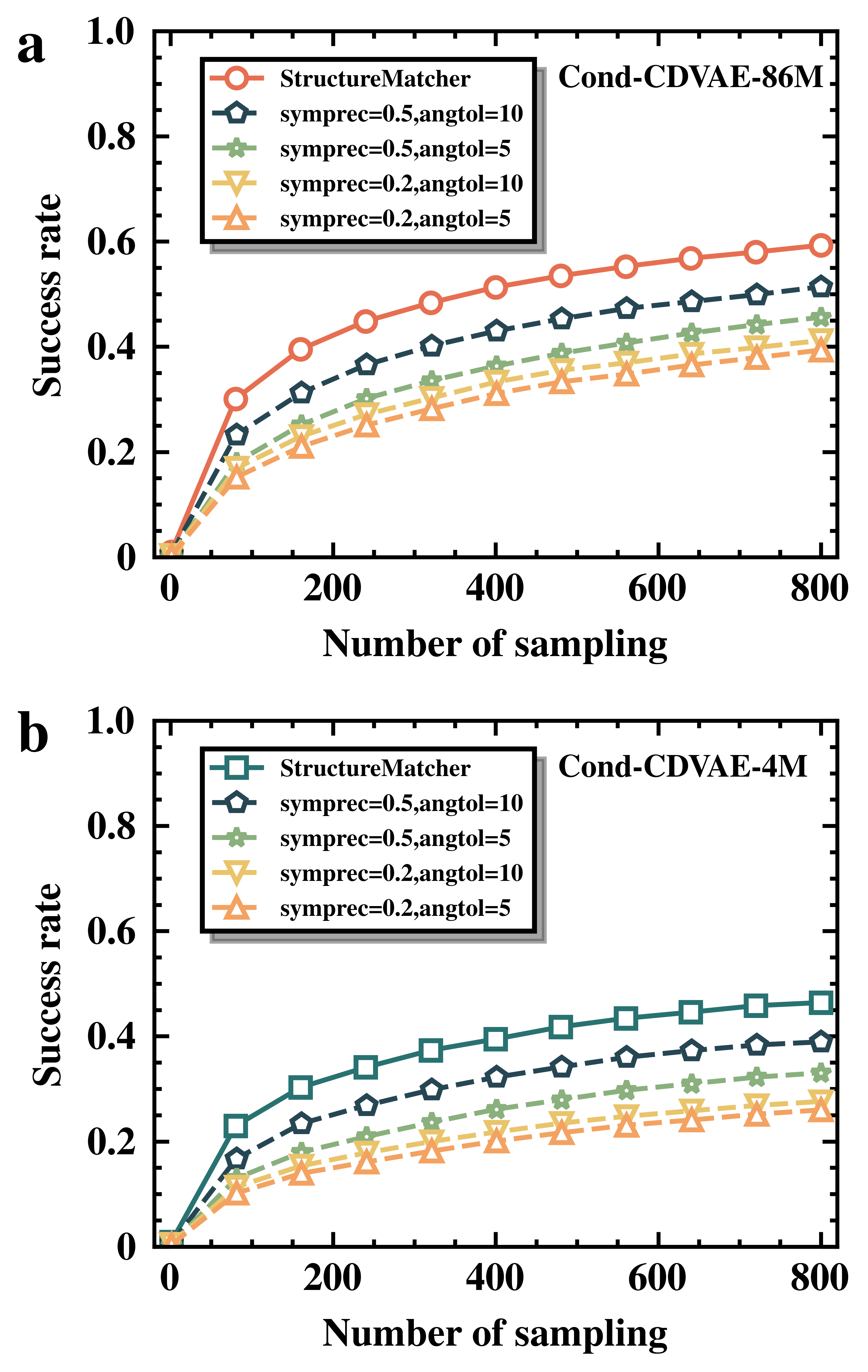}
\caption{\label{fig:sfig6} \textbf{The success rate of the Cond-CDVAE model on structure prediction of all 3,547 MP experimental structures in the test set as a function of number of structural samplings.} Results for models trained with \textbf{a} 86 and \textbf{b} 4 million parameters, respectively. Structure matching by solely using pymatgen \texttt{\detokenize{StructureMatcher}}, and in conjunction with space group determination by \texttt{\detokenize{SpacegroupAnalyzer}} of varying \texttt{\detokenize{symprec}} and \texttt{\detokenize{angle_tolerance}} parameters, are shown by solid and dashed lines, respectively.}
\end{figure}

\clearpage
\section*{Supplementary Tables}
\begin{table}[!ht]
\begin{threeparttable}
\caption{\label{tab:tab1}{Hyperparameters of Cond-CDVAE model used in this work} }
\begin{tabular}{
l
S[output-exponent-marker=e,exponent-product=,table-align-exponent]
S[output-exponent-marker=e,exponent-product=,table-align-exponent]
}
\toprule
    {} &
    {\fontseries{b}\selectfont \quad Cond-CDVAE-4M  \quad} &
    {\fontseries{b}\selectfont \quad Cond-CDVAE-86M \quad} \\
\midrule
    {\fontseries{b}\selectfont Model} \\
\cmidrule(r){1-1}
Element type embedding                                      & 50                & 50 \\  
Gaussian expansion for pressure, number of bases            & 80                & 80  \\  
Gaussian expansion for pressure, bases start                & -2.0              & -2.0 \\  
Gaussian expansion for pressure, bases stop                 & 5.0               & 5.0 \\  
{$\text{MLP}_{\text{zc}}$} number of layers                 & 3                 & 3   \\  
{$\text{MLP}_{\text{zc}}$} number of hidden channels        & 64                & 64  \\  
{$\text{MLP}_{\text{Lat}}$} number of layers                & 1                 & 1   \\  
{$\text{MLP}_{\text{Lat}}$} number of hidden channels       & 256               & 256 \\  
{$\text{PGNN}_{\text{Enc}}$} number of blocks               & 4                 & 8   \\  
{$\text{PGNN}_{\text{Enc}}$} number of hidden channels      & 128               & 512 \\  
{$\text{PGNN}_{\text{Enc}}$} interaction embedding size     & 128               & 256 \\  
{$\text{PGNN}_{\text{Dec}}$} number of blocks               & 4                 & 8   \\  
{$\text{PGNN}_{\text{Dec}}$} number ofhidden channels       & 128               & 512 \\  
Loss weight $\lambda_{\mathbf{L}}$                              & 10                & 10 \\  
Loss weight $\lambda_{\mathbf{X}}$                              & 10                & 10 \\  
Loss weight $\beta$                                         & 0.01              & 0.01 \\  
\toprule
    {\fontseries{b}\selectfont Optimizer} & &\\
\cmidrule(r){1-1}
Optimizer type                          & {Adam}                & {Adam} \\
Learning rate                           &  1e-4                 &  1e-4 \\
Learning rate scheduler                 & {ReduceLROnPlateau}   & {ReduceLROnPlateau} \\
Scheduler patience (epoch)              & 30                    & 30 \\
Scheduler factor (epoch)                & 0.6                   & 0.6 \\
Minimal learning rate                   & 1e-5                  & 1e-5 \\
\toprule
    {\fontseries{b}\selectfont Data} & & \\
\cmidrule(r){1-1}
Batch size & 128 & 128\\
\bottomrule
\end{tabular}
\end{threeparttable}
\end{table}

Each element type is embedded by a vector of length 50. Pressures are unit in GPa, and are first standardized on training set, then expanded to 80 equidistant Gaussian bases centered from -2.0 to 5.0 with standard deviation being the interval, i.e., $7/80$. $\text{MLP}_{\square}$ are fully connected layers. DimeNet++~\cite{Gasteiger.DimeNet.2020, Gasteiger.DimeNet++.2020} and GemNet-dQ~\cite{Gasteiger.GemNet.2021} are used as $\text{PGNN}_{\text{Enc}}$ and $\text{PGNN}_{\text{Dec}}$, respectively. A tanh activation layer is added at the end of $\text{PGNN}_{\text{Enc}}$ to improve the training stability.

\clearpage
\section*{Supplementary References}
% \bibliographystyle{apsrev4-2}
% \bibliographystyle{naturemag}
% \bibliography{ref}% Produces the bibliography via BibTeX.

%% file: table1.tex
\begin{table}
\begin{threeparttable}
\caption{\label{tab:tab1}\textbf{Statistic of the MP60-CALYPSO dataset.}}
\begin{tabular}{
l
S[table-format=6.0]
S[table-format=5.0]
S[table-format=2.0]
S[table-format=6.0]
}
    \toprule
        {} &
        {\fontseries{b}\selectfont structures}   &
        {\fontseries{b}\selectfont compositions} &
        {\fontseries{b}\selectfont elements}     &
        {\fontseries{b}\selectfont prototypes} \\
    \midrule
    MP60    &  99243 & 78227 & 85 &  32813 \\
    CALYPSO & 571736 &  7775 & 70 &  84643 \\
    Total   & 670979 & 85824 & 86 & 114733 \\
    \bottomrule
\end{tabular}
\end{threeparttable}

\end{table}

%% file: table2.tex
\begin{table}
\begin{threeparttable}
\caption{\label{tab:tab2}\textbf{Reconstruction performance of CDVAE and Cond-CDVAE on MP20 and MP60-CALYPSO datasets.} The match rate (MR) represents the percentage of structures successfully reconstructed in the held-out test set. The MP20 test set is consistent with that used in Ref. \cite{Xie.CDVAE.2022}, while the MP60-CALYPSO test set consists of 10\% of the entire dataset excluded from model training. The normalized root mean squared displacement (N-RMSD) is averaged across all matched structures.}
\begin{tabular}{
l
S[table-format=2.2]
S[table-format=1.4]
S[table-format=2.2]
S[table-format=1.4]
}
    \toprule
        \multirow[b]{2}{*}{\fontseries{b}\selectfont Method} &
        \multicolumn{2}{c}{\fontseries{b}\selectfont MP20} &
        \multicolumn{2}{c}{\fontseries{b}\selectfont MP60-CALYPSO} \\
\cmidrule(r){2-3} \cmidrule(r){4-5} &
        {\fontseries{b}\selectfont MR (\%)} &
        {\fontseries{b}\selectfont N-RMSD} &
        {\fontseries{b}\selectfont MR (\%)} &
        {\fontseries{b}\selectfont N-RMSD} \\
    \midrule
    CDVAE~\cite{Xie.CDVAE.2022}& 45.43 & 0.0356 & {-}   & {-}    \\
    Cond-CDVAE-4M              & 52.36 & 0.0767 & 13.01 & 0.2093 \\
    Cond-CDVAE-86M             & 57.50 & 0.0600 & 16.58 & 0.2066 \\
    \bottomrule
\end{tabular}
\end{threeparttable}
\end{table}

%% file: table3.tex
\begin{table}
\begin{threeparttable}
\caption{\label{tab:tab3}\textbf{Performance in DFT local optimizations of different types of structures.} Statistics are gathered from 500 structures of each type, including structures generated by the Cond-CDVAE-4M model, as well as random structures generated by CALYPSO, both with and without the constraints of space group symmetries.}
\begin{tabular}{
l
*{5}{S[table-format=2.2,detect-weight,mode=text]}
}
\toprule
\multicolumn{1}{r}{\fontseries{b}\selectfont P (GPa)\quad\quad\quad} & \fontseries{b}\selectfont 0 & \fontseries{b}\selectfont 100 & \fontseries{b}\selectfont 200 & \fontseries{b}\selectfont 300 & \fontseries{b}\selectfont 400 \\
\midrule
\multicolumn{1}{c}{\fontseries{b}\selectfont Convergence rate (\%) $\uparrow$}  \\
\cmidrule(r){1-1}
{Random without symmetry}    &    50.20 &    64.40 &    55.00 &    54.00 &    52.20 \\
{Random with symmetry}       &    79.80 &    67.80 &    60.40 &    54.00 &    49.80 \\
{Cond-CDVAE-4M}              & \fontseries{b}\selectfont 94.60 & \fontseries{b}\selectfont 97.00 & \fontseries{b}\selectfont 95.80 & \fontseries{b}\selectfont 91.40 & \fontseries{b}\selectfont 88.40 \\
\midrule
\multicolumn{1}{c}{\fontseries{b}\selectfont Average ionic step (\#) $\downarrow$}  \\
\cmidrule(r){1-1}
{Random without symmetry}    &    90.25 &    85.88 &    88.10 &    85.59 &    85.96 \\
{Random with symmetry}       &    50.81 & \fontseries{b}\selectfont 33.62 & \fontseries{b}\selectfont 31.41 & \fontseries{b}\selectfont 30.06 & \fontseries{b}\selectfont 30.70 \\
{Cond-CDVAE-4M}              & \fontseries{b}\selectfont 44.73 &    37.11 &    40.29 &    40.78 &    43.26 \\
\midrule
\multicolumn{1}{c}{\fontseries{b}\selectfont Average RMSD (\AA) $\downarrow$} \\         
\cmidrule(r){1-1}
{Random without symmetry}   &    1.88 &    1.39 &    1.36 &    1.29 &    1.34 \\
{Random with symmetry}       &    1.35 &    0.96 &    0.84 &    0.73 &    0.66 \\
{Cond-CDVAE-4M}              & \fontseries{b}\selectfont 0.79 & \fontseries{b}\selectfont 0.57 & \fontseries{b}\selectfont 0.54 & \fontseries{b}\selectfont 0.51 & \fontseries{b}\selectfont 0.48 \\
\bottomrule
\end{tabular}
\end{threeparttable}
\end{table}

%% file: table4.tex
\begin{table}
\begin{threeparttable}
\caption{\label{tab:tab4}\textbf{Performance comparison of structure prediction between the Cond-CDVAE-4M model and CALYPSO for typical high-pressure phases of Li, B, and \ce{SiO2}.} $N_\text{atoms}$ represents the number of atoms within a unit cell. $N_\text{model}$ and $N_\text{CSP}$ denote the average number of structural samplings required to locate the global minimum by the Cond-CDVAE-4M model and CALYPSO, respectively. Multiple independent simulations are conducted for each method, as indicated by Runs. When two numbers are provided, the first signifies the count of successful runs. A run is considered successful if it identifies the ground state within 1,000 samplings.}
\begin{tabular}{
l
S[table-format=2.0]
S[table-format=3.0]
S[table-format=3.1,detect-weight,mode=text]
c
S[table-format=3.1,detect-weight,mode=text]
c
}
    \toprule
        {\fontseries{b}\selectfont Structures} &
        {\fontseries{b}\selectfont $N_\text{atoms}$} &
        {\fontseries{b}\selectfont P~(GPa)} &
        {\fontseries{b}\selectfont $N_\text{model}$} &
        {\fontseries{b}\selectfont Runs} &
        {\fontseries{b}\selectfont $N_\text{CSP}$} &
        {\fontseries{b}\selectfont Runs} \\
    \midrule
 \ce{Li} \\
 \cmidrule(r){1-1}
 $cI16$                & 16 &   50 & 566.0     & 2/5    & \textbf{50.0} & \textbf{3/3} \\
     \midrule
 \ce{B} \\
 \cmidrule(r){1-1}
 {$\alpha$-\ce{B12}}       & 36 &    0 & \fontseries{b}\selectfont 74.0  & 1/5          & 392.3    & \textbf{3/3} \\
 {$\gamma$-\ce{B28}}       & 28 &   50 & \fontseries{b}\selectfont 341.0 & \textbf{5/5} &  {-}     &  0/3   \\
 {$\alpha$-Ga-type}        & 8  &  100 & \fontseries{b}\selectfont 58.0  & \textbf{5/5} &  78.0    &  3/3   \\
     \midrule
 \ce{SiO2} \\
 \cmidrule(r){1-1}
 {$\alpha$-quartz}         &  9 &    0 & \fontseries{b}\selectfont 62.8  & \textbf{5/5}  & 189.0      &  3/3   \\
 {coesite}                 & 24 &    5 & \fontseries{b}\selectfont 328.0 & \textbf{3/5}  &  {-}       &  0/3   \\
 {rutile-type}             &  6 &   50 & \fontseries{b}\selectfont 5.0   & \textbf{5/5}  &  90.0      &  3/3   \\
 {\ce{CaCl2}-type}         &  6 &   80 & \fontseries{b}\selectfont 10.2  & \textbf{5/5}  &  61.0      &  3/3   \\
 {$\alpha$-\ce{PbO2}-type} & 12 &  100 & \fontseries{b}\selectfont 21.6  & \textbf{5/5}  &  74.7      &  3/3   \\
 {pyrite-type}             & 12 &  300 & \fontseries{b}\selectfont 28.8  & \textbf{5/5}  &  66.0      &  3/3   \\
    \bottomrule
\end{tabular}
\end{threeparttable}
\end{table}